\documentclass[conference]{IEEEtran}
\IEEEoverridecommandlockouts
\usepackage{cite}
\usepackage{amsmath,amssymb,amsfonts}
\usepackage{algorithmic}
\usepackage{graphicx}
\usepackage{textcomp}
\usepackage{xcolor}
\usepackage{float}
\usepackage{tabularx} 
\usepackage{subcaption}
\usepackage[hidelinks]{hyperref}
\usepackage[linesnumbered,ruled,vlined]{algorithm2e}
\def\BibTeX{{\rm B\kern-.05em{\sc i\kern-.025em b}\kern-.08em
    T\kern-.1667em\lower.7ex\hbox{E}\kern-.125emX}}

\makeatletter
\def\ps@IEEEtitlepagestyle{%
  \def\@oddhead{}% clear header
  \def\@evenhead{}%
  \def\@oddfoot{\mycopyrightnotice}% place notice in footer
  \def\@evenfoot{}%
}

\def\mycopyrightnotice{%
  \begin{minipage}{\textwidth}
    \centering
    \footnotesize
    © 2025 IEEE. Personal use of this material is permitted. Permission from IEEE must be obtained 
    for all other uses, in any current or future media, including reprinting/republishing 
    this material for advertising or promotional purposes, creating new collective works, 
    for resale or redistribution to servers or lists, or reuse of any copyrighted component 
    of this work in other works.
  \end{minipage}%
  \gdef\mycopyrightnotice{}% ensure it only appears once
}
\makeatother

\begin{document}

\title{Constrained Network Adversarial Attacks: Validity, Robustness, and Transferability
}

\thispagestyle{IEEEtitlepagestyle}
% \author{\IEEEauthorblockN{Oumaima Taheri}
% \IEEEauthorblockA{\textit{Ai movement} \\
% \textit{The International Artificial Intelligence}\\ \textit{Center of Morocco,}
% \textit{UM6P}\\
% Rabat, Morocco \\
% Oumaima.TAHERI@um6p.ma}
% \and
% \IEEEauthorblockN{Btissam El Khamlichi}
% \IEEEauthorblockA{\textit{Ai movement} \\
% \textit{The International Artificial Intelligence}\\ \textit{Center of Morocco,}
% \textit{UM6P}\\
% Rabat, Morocco \\
% Btissam.ELKHAMLICHI@um6p.ma}
% \and
% \IEEEauthorblockN{ Amal El Fallah Seghrouchni}
% \IEEEauthorblockA{\textit{Ai movement} \\
% \textit{The International Artificial Intelligence}\\ \textit{Center of Morocco,}
% \textit{UM6P}\\
% Rabat, Morocco \\
% Amal.elfallah-seghrouchni@um6p.ma}

% }

\author{
    \vspace{3mm}
    \IEEEauthorblockN{\textbf{Anass Grini}\IEEEauthorrefmark{1},
                      \textbf{Oumaima Taheri}\IEEEauthorrefmark{1},
                      \textbf{Btissam El Khamlichi}\IEEEauthorrefmark{1},
                      \textbf{Amal El Fallah-Seghrouchni}\IEEEauthorrefmark{1}\IEEEauthorrefmark{4}}
    % \vspace{3mm}
    
    \IEEEauthorblockA{\IEEEauthorrefmark{1}Ai movement, The International Artificial Intelligence Center of Morocco, UM6P, Rabat, Morocco}
    \IEEEauthorblockA{\IEEEauthorrefmark{4}Sorbonne University, LIP6 - UMR 7606 CNRS, Paris, France}

    \{anass.grini, btissam.elkhamlichi, amal.ElFallah-Seghrouchni\}@um6p.ma
}

\maketitle

\begin{abstract}

While machine learning has significantly advanced Network Intrusion Detection Systems (NIDS), particularly within IoT environments where devices generate large volumes of data and are increasingly susceptible to cyber threats, these models remain vulnerable to adversarial attacks. Our research reveals a critical flaw in existing adversarial attack methodologies: the frequent violation of domain-specific constraints, such as numerical and categorical limits, inherent to IoT and network traffic. This leads to up to 80.3\% of adversarial examples being invalid, significantly overstating real-world vulnerabilities. These invalid examples, though effective in fooling models, do not represent feasible attacks within practical IoT deployments. Consequently, relying on these results can mislead resource allocation for defense, inflating the perceived susceptibility of IoT-enabled NIDS models to adversarial manipulation. Furthermore, we demonstrate that simpler surrogate models like Multi-Layer Perceptron (MLP) generate more valid adversarial examples compared to complex architectures such as CNNs and LSTMs. Using the MLP as a surrogate, we analyze the transferability of adversarial severity to other ML/DL models commonly used in IoT contexts. This work underscores the importance of considering both domain constraints and model architecture when evaluating and designing robust ML/DL models for security-critical IoT and network applications.

\end{abstract}

\begin{IEEEkeywords}
Network Intrusion Detection, Internet of Things, Adversarial attacks, Cybersecurity, Machine Learning
\end{IEEEkeywords}

\section{Introduction}
ML has become a transformative technology across numerous domains, including the rapidly growing area of the Internet of Things (IoT). In cybersecurity, particularly within IoT environments, ML demonstrates significant proficiency in accurately classifying network traffic as legitimate or malicious \cite{andresini2021insomnia}. However, the integration of ML into IoT introduces vulnerabilities, as research indicates ML-based Network Intrusion Detection Systems (NIDS) in IoT are susceptible to adversarial examples—inputs deliberately crafted to induce misclassification \cite{alhajjar2021adversarial}.

The field of Adversarial Machine Learning (AML) originated in computer vision, where minor, often imperceptible alterations to images can deceive classifiers \cite{szegedy2013intriguing}. These attacks exploit vulnerabilities in pre-trained models by subtly shifting data points across decision boundaries to produce incorrect outputs. ML algorithms' inherent pattern-recognition capabilities make them particularly vulnerable to evasion by adversaries seeking to bypass detection, especially in IoT networks characterized by diverse and resource-constrained devices.

Traditionally, AML research emphasizes unconstrained domains, offering attackers full control over the feature space. This scenario is unrealistic for IoT and cybersecurity applications, where adversaries face strict domain-specific constraints and limited control over feature manipulation \cite{anthi2021adversarial}. Unlike image data, IoT-generated network data is substantially harder to alter without breaking functional dependencies intrinsic to network protocols and IoT device operations.

Although considerable research has been conducted on improving ML model robustness \cite{xiao2021ebsnn,grini2024hpac}, insufficient attention has been paid to the fundamental constraints inherent in IoT and network security domains. These domain-specific constraints present unique challenges, such as:

\begin{itemize}
\item \textbf{Perturbation Imperceptibility}: IoT network data does not involve human perception, making human-imperceptible perturbations irrelevant.
\item \textbf{Feature Space Control}: IoT features reflect specific network behaviors, severely limiting an adversary's capacity to freely alter them.
\end{itemize}

These challenges underscore the necessity for targeted AML approaches tailored to IoT contexts. This paper provides a comprehensive evaluation of AML techniques, explicitly accounting for IoT and network security constraints. We investigate state-of-the-art AML attacks' effectiveness on NIDS under realistic constraints and evaluate adversarial attacks' transferability across various IoT models, particularly under limited attacker knowledge (black-box conditions).

Our research further identifies a critical relationship between model complexity and robustness against adversarial attacks within IoT and network security domains. This finding emphasizes that choosing the right surrogate model significantly impacts the generation and effectiveness of adversarial examples within constrained environments.

Specifically, this work contributes by:

\begin{itemize}
\item[-] Formalizing and implementing domain constraints in an IoT network traffic scenario.
\item[-] Proposing a novel validation process for adversarial examples that incorporates \textit{Numerical Dependencies} and \textit{Categorical Dependencies} to ensure compliance with IoT and network constraints.
\item[-] Measuring the validity of adversarial examples generated by existing algorithms by projecting them onto a feasible input space.
\item[-] Analyzing how model complexity influences the generation of valid adversarial examples, highlighting increased robustness in sophisticated IoT model architectures.
\item[-] Evaluating the effectiveness and transferability of feasible adversarial examples within constrained IoT network environments.
\end{itemize}

The remainder of this paper is structured as follows: Section \ref{section:2} reviews relevant literature on adversarial example generation in constrained IoT domains. Section \ref{section:3} details our methodology for modeling constraints, specifically defining the feasible adversarial example space. Section \ref{section:4} presents our experimental setup, results, and discusses their implications. Finally, Section \ref{section:5} concludes the paper and suggests directions for future research.

%============================================================================================================================

% \begin{figure}[ht!] %!t
% \includegraphics[width=7cm, height=4cm]{figure paper.PNG}
% \caption{\smallAdversarial attack is a mapping M such that the perturbed instance $M(X_m)$ is misclassified as $C_t$.}
% \label{decision}
% \end{figure}

\section{Related work}\label{section:2}
Since their formal introduction in 2013 \cite{szegedy2013intriguing}, AML techniques have posed a growing risk across various domains. While early research on AML emphasized image classification, cybersecurity presents unique challenges due to its use of categorical, continuous, and discrete feature types.

Although research interest in this area is growing, some studies overlook real-world constraints, which can limit the practical applicability of their AE generation techniques. The current literature exhibits several limitations:
\begin{itemize}
\item \textit{Lack of realistic threat modeling:} Works by \cite{apruzzese2019addressing}, \cite{martins2020adversarial}, and \cite{de2019adversarial} showcase the potential of AML for cyberattacks, yet they do not explicitly address the real-world constraints present in actual network environments.

\item \textit{Arbitrary network traffic generation:} Yang et al. \cite{yang2018adversarial} utilize arbitrary network traffic without establishing realistic perturbation limits, while studies like \cite{rigaki2017adversarial} and \cite{lin2018idsgan} often overlook or bypass important domain-specific constraints.

\end{itemize}

Recent works addressing IoT-specific challenges include Kumar et al. \cite{kumar2024nidscbad}, who proposed NIDS-CBAD, a constraint-based adversarial detection method tailored to IoT networks, significantly enhancing detection efficiency. Additionally, Sharma and Chen \cite{sharma2024systematic} systematically analyzed adversarial attacks on ML-based NIDS in IoT environments, highlighting varying vulnerabilities across ML models.

While recent works in AML for cybersecurity acknowledge the importance of network-specific constraints, they often lack detailed consideration of these constraints. Venturi \cite{venturi2021feasibility} discusses adversarial solutions without addressing specific network constraints. Hashemi et al. \cite{hashemi2019towards} focus on feature dependencies and limited attacker control in flow-based NIDS but do not fully explore all network constraints. Teuffenbach et al. \cite{teuffenbach2020subverting} enhance the Carlini \& Wagner attack by incorporating some feature limitations and weightings, yet they do not cover the complete range of network-specific constraints. These studies highlight significant progress but also reveal gaps in comprehensively addressing all network-specific constraints in AML.

Sheatsley et al. \cite{sheatsley2020adversarial} build upon investigations into adversarial examples in constrained domains. While it provides a valuable exploration of categorical constraints, such as protocol and service dependencies, it gives less attention to numerical dependencies. Features like duration, packet sizes, and byte counts, which can carry important continuous relationships in network data, are not as prominently addressed.

%=================================================================================================

\section{Methodology}\label{section:3}
% This work aims to assess the validity of adversarial examples generated by existing algorithms against ML and DL-based NIDS. Unlike previous studies, we explicitly model and analyze domain constraints, including \textit{numerical} and \textit{categorical dependencies} in feature space, based on domain knowledge and data observations
% , as shown in Figure \ref{picture}.

This work is focused on the realistic feasibility of the proposed threats against ML and DL-based Network Intrusion Detection Systems (NIDS). It aims to validate adversarial examples generated by existing algorithms within the context of domain constraints. Unlike previous studies, we provide a comprehensive model and analysis of these constraints, including numerical and categorical dependencies in feature space derived from domain knowledge and data observations.

% \subsection{General Description}
% The threat model in networks is unique in that there are domain limitations that adversaries must follow to craft a valid traffic flow. We introduce the concept of primary features. This key feature, by definition, is one that, when set to a specific value, restricts the range of allowable values for other features. Many popular network intrusion detection datasets include services, flags, and additional protocol-related information that can be directly used as ML features. The concept of primary features is particularly well suited to networks. When creating adversarial instances, constraints encode the maneuvers that are conceivable for an adversary. We describe two main dependencies in feature space: Categorical and numerical Dependencies, as shown in Figure \ref{picture}.

% \begin{figure*}[] %!t
% \centering
% \includegraphics[width=4.4in]{schema of work.pdf}
% \caption{\small A general architecture of constraints extraction}
% \label{picture}
% \end{figure*}

\subsection{Categorical  Dependencies}
The adversary must follow the TCP/IP protocol to carry out network attacks. Any feature vector that breaks the TCP/IP protocol is not valid. These dependencies are defined using \textit{primary features} identified through understanding the domain and data observation \cite{sheatsley2020adversarial}. Primary features are, by definition, substantially correlated with the majority of features, meaning that changing one of them will affect a subset of the others, which must be updated correspondingly. Because most ML-based NIDS features are related to protocols, it is intuitive to have transport layer protocols as primary features.

Messages traveling a network without set rules or procedures would be unformatted and may not be intelligible to the receiving device. Protocols will be the conditionals on the values of other secondary features, such as flags and services. The following is a description of this relationship \cite{sheatsley2020adversarial}:
\[
   \forall \mathbf{x}\in \mathbb{X}: \mathbf{x}_p\Rightarrow (\mathbf{x}_1\in \mathbb{Y}_1)\wedge(\mathbf{x}_2\in \mathbb{Y}_2)\wedge ... \wedge (\mathbf{x}_n\in \mathbb{Y}_n) 
\]

Where $\mathbf{x}$ is an input, $\mathbb{X}$ is the dataset, $\mathbf{x}_p$ is the key feature, and $\mathbb{Y}_i$ indicates the range of values that the semantics of feature $\mathbf{x}_{i\in [1,n]}$  allow (where $n$ is the number of features). A network limitation between TCP/UDP and service type can be represented as follows.
\[
   \forall \mathbf{x}\in \mathbb{X}: \mathbf{x}_{TCP}\Rightarrow \mathbf{x}_{service}\in [nntp,ssh,ftp \_ data,smtp,...]
\]
\[
   \forall \mathbf{x}\in \mathbb{X}: \mathbf{x}_{UDP}\Rightarrow \mathbf{x}_{service}\in [bootps,tftp,ntp,snmp,...]
\]

\subsection{Numerical Dependencies}
While adversarial attacks often consider categorical dependencies, numerical dependencies remain underexplored. In this paper, we address these numerical dependencies, which can naturally emerge in raw data or through feature engineering. Examples include one-hot encoded features such as service type, connection type, and flags, as well as continuous values like duration (the time between the first and last packet), original bytes, received bytes, and packet counts. By incorporating both categorical and numerical dependencies, we aim to provide a more comprehensive approach to adversarial attack generation within constrained domains.

To ensure the validity and realism of the adversarial examples, we apply a filtering process that enforces both categorical and numerical constraints, as described in Algorithm \ref{alg:tcp_dependency}. Specifically, binary features in the dataset are rounded to their nearest integer values; if a perturbation results in a binary feature taking on a value like $1+\epsilon$, it is rounded back to $1$, and similarly for values near $0$. This ensures that binary features remain within their valid set of $\{0,1\}$. Continuous (floating-point) features are where adversarial perturbations have an effect, enhancing the realism of the generated examples.

In addition to the numerical dependencies we've addressed, the connection type exhibits one-hot encoded (OHE) categorical dependencies that must be respected. Specifically, for a set of binary variables $\{ \mathbf{x}_{i=1}^{N}   \}$, where each $\mathbf{x} \in \{0,1\}$ and $N$ is the total number of unique categories, the following condition holds:
\[
\sum_{i=1}^{N} x_i = 1
\]
Here, $x_i$ represents a categorical feature. For example, consider the feature \textit{"protocol\_type"}, which has categorical values "tcp", "udp", and "icmp". Using one-hot encoding, we create three new binary features: \textit{"protocol\_type\_tcp"}, \textit{"protocol\_type\_udp"}, and \textit{"protocol\_type\_icmp"}. If the protocol type is TCP, then \textit{"protocol\_type\_tcp"} is set to 1, and both \textit{"protocol\_type\_udp"} and \textit{"protocol\_type\_icmp"} are set to 0. This same process applies to other categorical features like services and flags.
To ensure data integrity, any adversarial example that violates this OHE restriction is filtered out. Additionally, binary fields impose further limits on data modification; when modifying binary values, only 0 or 1 can be used as the changed value.

Algorithm \ref{alg:tcp_dependency} acts as a filter for adversarial examples, retaining only valid samples that ensure both numerical and categorical dependencies for the TCP case.

\begin{algorithm}[h]
\label{alg:tcp_dependency}
\caption{Filter Adversarial Examples for Categorical and Numerical Dependencies in TCP Case}
\small
\SetAlgoLined
\KwIn{- $x_{adv}$: Adversarial Examples (structured as a dataframe), \newline 
       - Service\_tcp, Service\_udp, Service\_icmp: Lists of protocol-specific services, \newline
       - Flag\_tcp, Flag\_udp, Flag\_icmp: Lists of protocol-specific flags, \newline
       - Binary\_features: List of features expected to be binary}
\KwOut{Valid adversarial examples $x_{valid}$}

\textbf{Convert} $x_{adv}$ to dataframe $adversarial\_data$\;
\textbf{Initialize} $x_{valid} \gets []$\;

\For{$i = 1$ \textbf{to} size($x_{adv}$)[0]}{
    Cast binary features and protocol type fields of $adversarial\_data_i$ to integer\;
    \If{$adversarial\_data_i[protocol\_type\_tcp] == 1$}{
        Ensure $adversarial\_data_i[protocol\_type\_udp] = 0$ and $adversarial\_data_i[protocol\_type\_icmp] = 0$\;
        \For{each $feature$ in Binary\_features}{
            Ensure $adversarial\_data_i[feature] \in \{0, 1\}$\;
        }
        \If{any($adversarial\_data_i[service] == 1$) for $service \in Service\_tcp$}{
            \If{any($adversarial\_data_i[flag] == 1$) for $flag \in Flag\_tcp$}{
                Append $adversarial\_data_i$ to $x_{valid}$\;
            }
        }
    }
}
\Return $x_{valid}$\;

\end{algorithm}
% \vspace{-.15cm}

% \begin{algorithm}[H]
% \DontPrintSemicolon
% \caption{Extract from Categorical Dependency Algorithm: TCP Case}
% \label{alg:tcp_dependency}

% \textbf{Require:} 
% \begin{itemize}
% \small
% \item[-] Service\_tcp: list of TCP services 
% \item[-] Service\_udp: list of UDP services
% \item[-] Service\_icmp: list of ICMP services
% \item[-] Flag\_tcp: list of TCP flags
% \item[-] Flag\_udp: list of UDP flags
% \item[-] Flag\_icmp: list of ICMP flags
% \item[-] Binary\_features: list of binary attributes
% \end{itemize}

% \textbf{Input:}  $x_{adv}$ (adversarial examples)

% \textbf{Output:} List of valid adversarial examples

% \begin{algorithmic}[1]
% \small
% \State Convert $x_{adv}$ to dataframe adversarial\_data;
% \For {i = 0 to size($x_{adv}$)[0]} 
% \If {$adversarial\_data_{i}$$[protocol\_type:tcp]$ = 1} 
    
%         \State \textbf{Ensure:} 
%             \begin{itemize}
%                 \item $adversarial\_data_{i}$$[protocol\_type:udp]$ = 0 
%                 \item $adversarial\_data_{i}$$[protocol\_type:icmp]$ = 0 
%                 \item binary\_features $\in$ [0, 1] 
%             \end{itemize}
%          \For{s in Service\_tcp} 
%             \If{$adversarial\_data_{i}$[s] = 1}
            
%                 \For{f in Flag\_tcp}
                
%                     \If{$adversarial\_data_{i}$[f] = 1}
                    
%                         \State valid\_adversarial $\gets$ $adversarial\_data_i$ ;
                        
%                     \EndIf
%                 \EndFor 
%             \EndIf 
%         \EndFor 
%     \EndIf 
% \EndFor
% \State \textbf{return} valid\_adversarial 
% \end{algorithmic}
% \end{algorithm}

\section{Experimental Results}\label{section:4}
\subsection{Evaluation Setting}
% In this section, we present the results of  experimenting our approach on the network intrusion dataset NSL-KDD. Subsection (A) describes the dataset and the generic adversarial attacks used against the five ML-Based IDSs. In subsection(B), we examine the robustness of models against these attacks. In (C), we mesure the validity of adversarial examples. Finally, the impact of adversarial attacks' transferability property are illustrated in subsection (D).

\subsubsection{\textbf{Dataset}}

We utilize the established NSL-KDD dataset \cite{dhanabal2015study} to evaluate our approach, as it offers a balanced and diverse range of network traffic features, including categorical, binary, discrete, and continuous types. To prepare data, we employ standard preprocessing:  one-hot encoding for categorical features and min-max normalization for overall stability and bias reduction. Importantly, we generate adversarial examples solely from the malicious traffic within NSL-KDD. This isolates the impact of adversarial techniques on misclassification, allowing us to directly assess how ML/DL-based NIDS models are affected within the specific context of intrusion detection.

\subsubsection{\textbf{Attacks}}

% In order to generate adversarial examples, we use seven state of the art adversarial attack algorithms: Fast Gradient Sign Method (FGSM) \cite{goodfellow2014explaining},  Carlini \& Wanger (C\&W) \cite{carlini2017towards}, Jacobian-Based Saliency Map Attack (JSMA) \cite{papernot2016limitations}, DeepFool \cite{moosavi2016deepfool}, Projected Gradient Decent (PGD) \cite{madry2017towards} , Zeroth order optimization (ZOO) \cite{chen2017zoo} and Basic Itertaive Method (BIM) \cite{kurakin2016adversarial}.

We use a suite of seven well-established adversarial attack algorithms for comprehensive evaluation: Fast Gradient Sign Method (FGSM) \cite{goodfellow2014explaining}, Carlini \& Wanger (C\&W) \cite{carlini2017towards}, Jacobian-Based Saliency Map Attack (JSMA) \cite{papernot2016limitations}, DeepFool \cite{moosavi2016deepfool}, Projected Gradient Descent (PGD) \cite{madry2017towards}, Zeroth Order Optimization (ZOO) \cite{chen2017zoo}, and Basic Iterative Method (BIM) \cite{kurakin2016adversarial}. These techniques represent a diverse range of strategies for crafting adversarial perturbations.

First, we consider a multi-layer perception (MLP) pre-trained on NSL-KDD dataset as a target model for adversarial attacks, encompassing both benign and malicious traffic samples.  The MLP architecture consists of three hidden layers with 512, 256, and 64 units respectively, utilizing ReLU activations. To mitigate overfitting, dropout regularization is incorporated after each hidden layer (rate of 0.01).  The output layer employs a Sigmoid activation, with Adam optimization and cross-entropy loss for the training process.

\subsection{Attacks feasibility}
% Following the implementation of domain restrictions, we project adversarial instances generated by existing algorithms onto the space of valid inputs to determine their viability. The percentage of non feasible adversarial examples created by each attack technique is shown in figure \ref{feasibility}.

\begin{figure}[h!] %!t
\includegraphics[width=9cm, height=5.6cm]{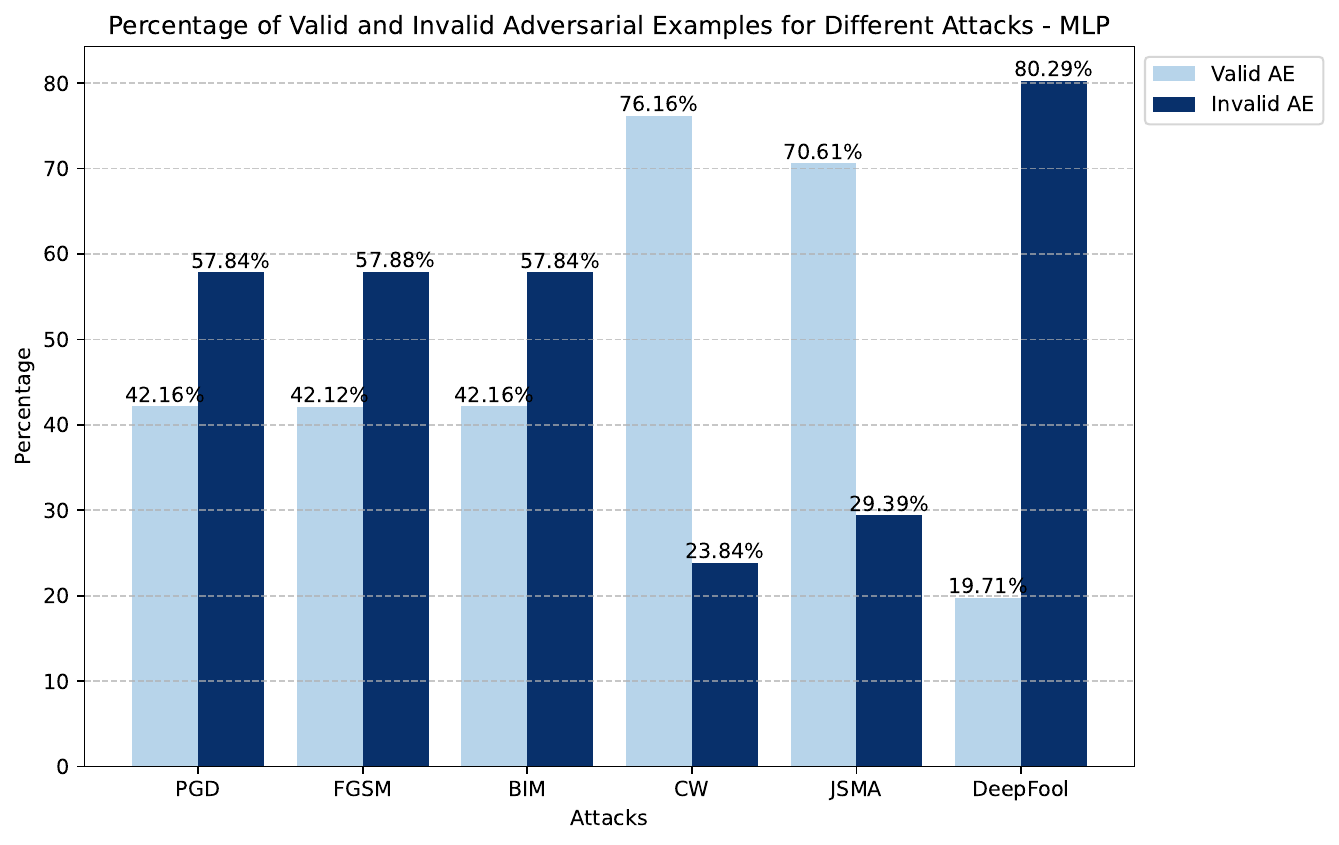}
\caption{\small Validity Percentage of AEs for different attacks.}
\label{feasibility}
\end{figure}

Upon implementing domain constraints, the feasibility of existing adversarial instances is rigorously assessed through their projection onto the space of valid inputs. Adversarial instances deemed \textit{invalid} are represented by samples prior to the application of any filtering mechanism. Figure \ref{feasibility} shows the proportion of valid and invalid adversarial examples produced by each attack method, highlighting notable differences.

% The findings of the validity evaluation of adversarial examples reveal that these attack methods do not adhere to domain limitations. Constraints that are defined by the following dependencies: numerical dependencies (the values within a feature may be fixed) and categorical dependencies (the values of different features may be correlated). Only few allowable inputs are crafted by exisiting approaches. According to these findings, existing algorithms can create up to 80.3\% invalid adversarial instances. As a result, the assessment based on these attacks does not reflect the true vulnerabilities. JSMA would be the most pratical in real scenarios (33,63\% unfeasible AEs) as it perturbs a small range of features, it only picks and changes the most likely attributes that makes that largest increase (largest gradient). FGSM, PGD, and BIM are ineffective methods for evading ML-based NIDS since they change 100\% of the traffic attributes, resulting in a significant percentage of invalid adversarial instances. The percentage of unfeasible adversarial examples in C\&W attack is 18.86\%, this attack provides data points that are quite similar to valid samples, but are intentionally engineered to cause minimum distortion and drive the network to make the wrong decision.

The assessment of adversarial example validity reveals a critical weakness in existing attack methodologies: a disregard for domain-specific constraints. In particular, numerical constraints (e.g., bounded feature values) and categorical constraints (e.g., inter-feature relationships) are frequently violated. This leads to a significant proportion of generated adversarial instances being \textit{invalid}, with our findings indicating rates as high as 80.3\%. Consequently, evaluations based on such attacks may substantially overstate real-world vulnerabilities. The discrepancy stems from the fact that these adversarial examples, while effective in deceiving the model, do not constitute plausible inputs that an attacker could realistically generate in practice. As such, relying on these results could lead to an inflated perception of the model's susceptibility to adversarial manipulation, potentially misguiding resource allocation for defensive measures.

Among the evaluated attacks, JSMA exhibits the highest real-world practicality due to its targeted modification of a limited feature set. Conversely, PGD, FGSM, and BIM are less suitable for attacking ML-based NIDS given their propensity to modify all traffic attributes. While C\&W yields the lowest rate of invalid adversarial instances, its effectiveness hinges on crafting minimally distorted data points closely resembling valid samples, leading to successfully engineered misclassifications.

% \vspace{-0.4cm}

For a good integration of ML in IDS, it's crucial to determine the resilience or susceptibility of constrained environments to adversarial attacks. By restricting adversarial examples to the domain of permissible inputs, we conduct a more rigorous and realistic evaluation of attack effectiveness.

% \begin{table}[ht]
% \centering
% \caption{Results of feasible adversarial attacks on MLP}
% \label{table3}
% \begin{tabular}[t]{lccccccc}
% \multicolumn{7}{c}{ACCURACY}\\
% \hline
% &JSMA&FGSM&DeepFool&C\&W&PGD&BIM\\
% \hline
% MLP& 5.12 & 98.02 & 100 & 99.51 & 100 & 100\\

% \hline
% \end{tabular}
% \end{table}%

% The performance of only feasible adversarial examples is significantly better than all inputs crafted by attack algorithms in terms of accuracy.
% The JSMA reduces the accuracy of the baseline model MLP, which is used to produce adversarial samples on the test set, by 54.22\%. ( from 79,11 to 36,21). MLP's performance was successfully degraded by the FGSM attack, which resulted in a 72.92\% decline in accuracy. With DeepFool, the accuracy drops from 79.11\% to 38.70\%. The C\&W and ZOO attack have a similar impact, resulting in a 54\% reduction in accuracy. The results show that using the Basic Iteraive approach reduces accuracy by 66,74\%.
% The Projected Gradient Descent (PGD) attack can reach a drop of accuracy of 81,94\% with valid instances. 

\begin{figure}[h!] %!t
\includegraphics[width=9cm, height=6.2cm]{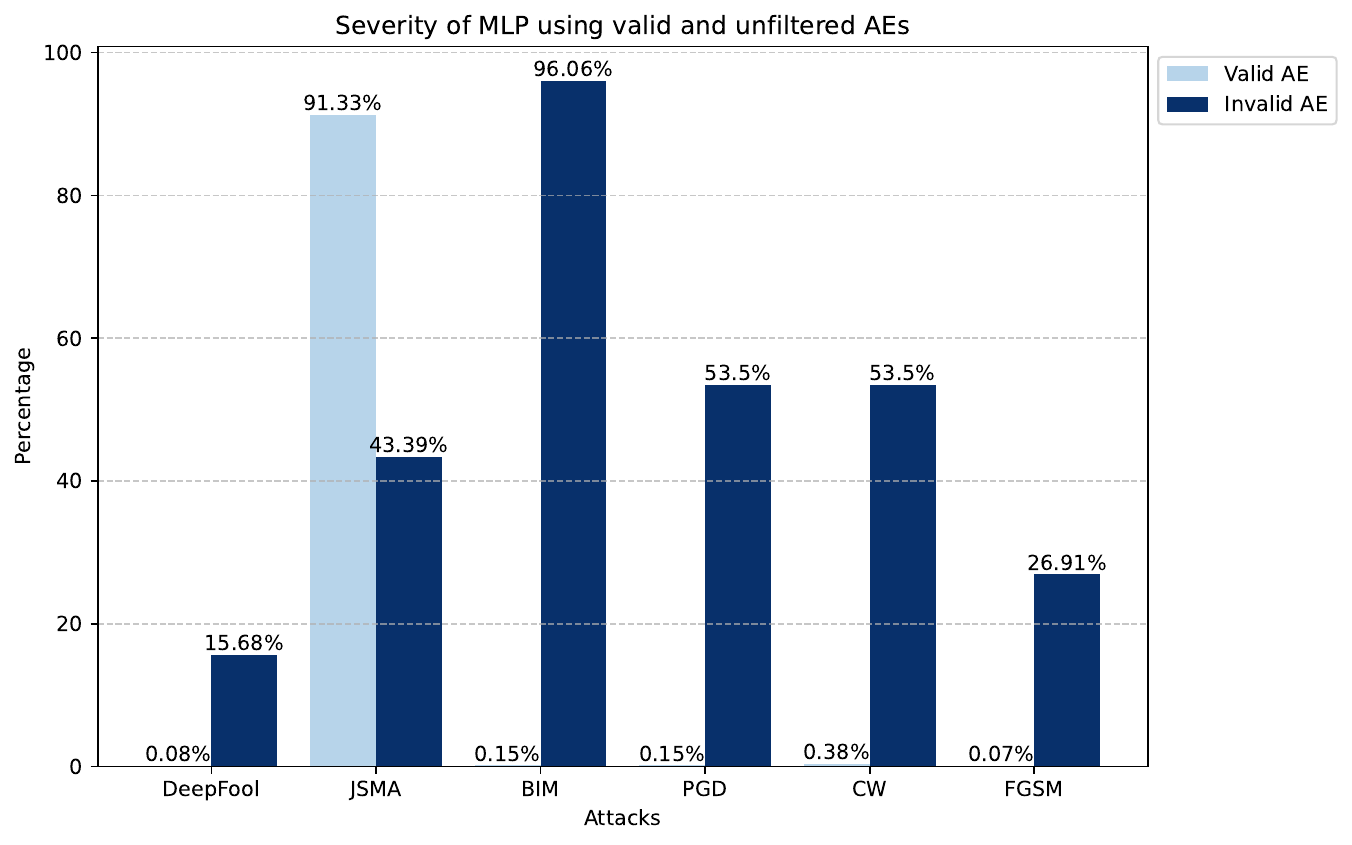}
\caption{\small Vulnerability assessment results with Valid and Invalid Adversarial Instances.}
\label{MLPVul}
\end{figure}

Imposing domain constraints on adversarial examples significantly reduces classifier vulnerability, with severity rates decreasing by 52.49\% to 99.84\% across most attacks (Figure \ref{MLPVul}). This substantial reduction indicates that many adversarial perturbations become less effective when restricted to feasible inputs.
\begin{figure*}[t!]
        \centering
        \begin{subfigure}[b]{0.435\textwidth}
            \centering
            \includegraphics[width=8.5cm, height=6cm]{img/new_validity_MLP.pdf}
            \caption[MLP as surrogate model]%
            {{\small MLP as surrogate model}}    
            \label{fig:mean and std of net14}
        \end{subfigure}
        \hfill
        \begin{subfigure}[b]{0.455\textwidth}  
            \centering 
            \includegraphics[width=8.5cm, height=6cm]{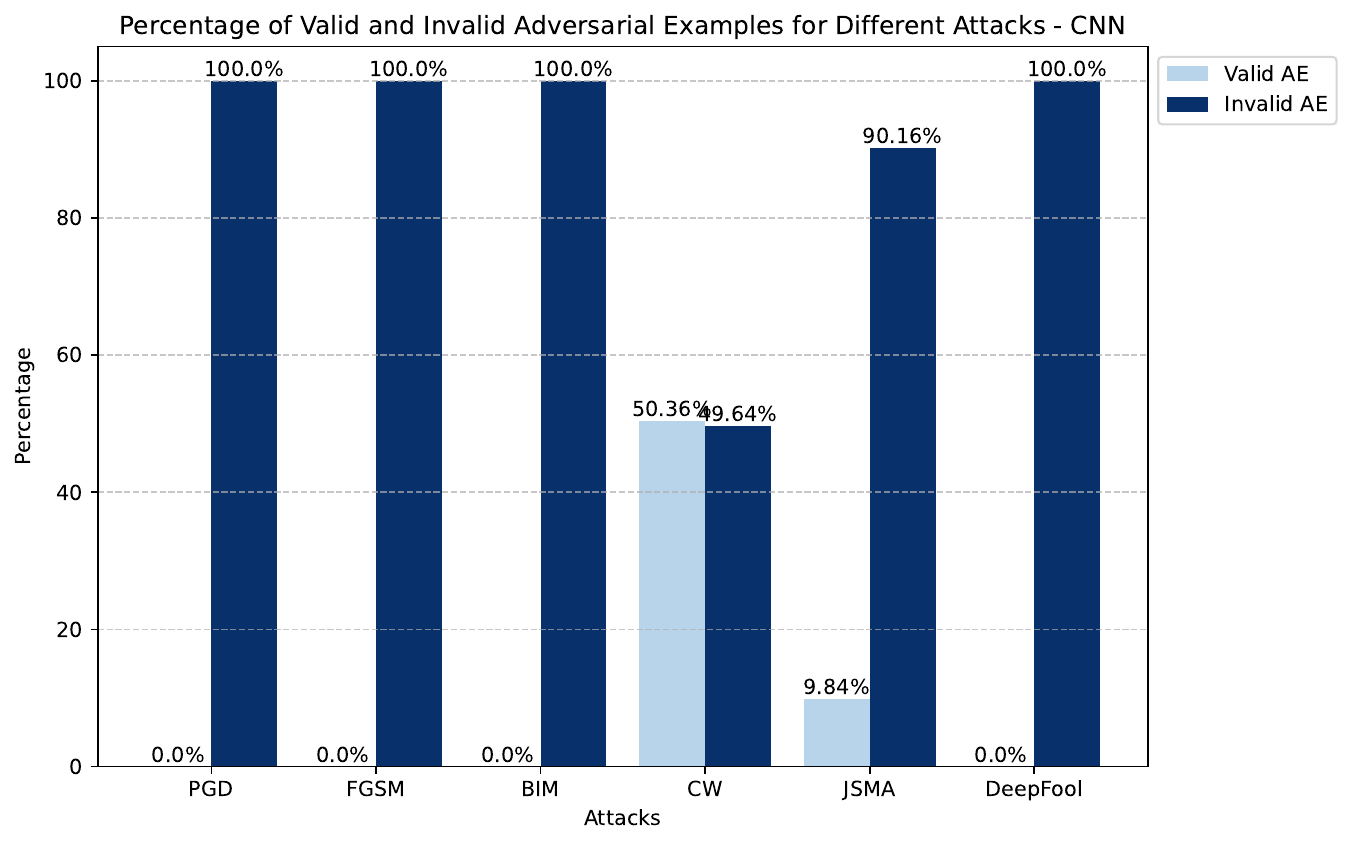}
            \caption[CNN as surrogate model]%
            {{\small CNN as surrogate model}}    
            \label{fig:mean and std of net24}
        \end{subfigure}
        \vskip\baselineskip
        \begin{subfigure}[b]{0.455\textwidth}   
            \centering 
            \includegraphics[width=8.5cm, height=6cm]{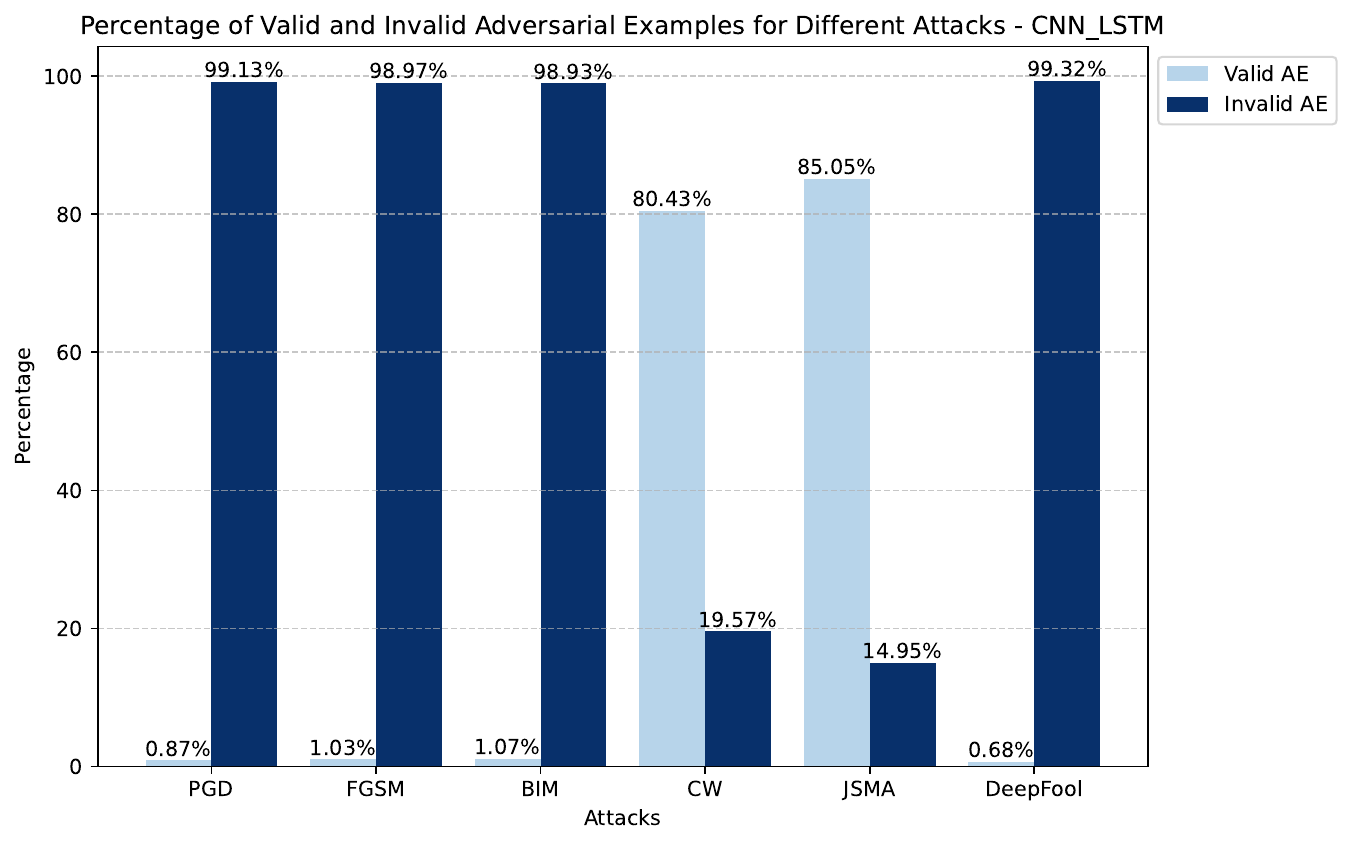}
            \caption[]%
            {{\small CNN-LSTM as surrogate model}}    
            \label{fig:mean and std of net34}
        \end{subfigure}
        \hfill
        \begin{subfigure}[b]{0.455\textwidth}   
            \centering 
            \includegraphics[width=8.5cm, height=6cm]{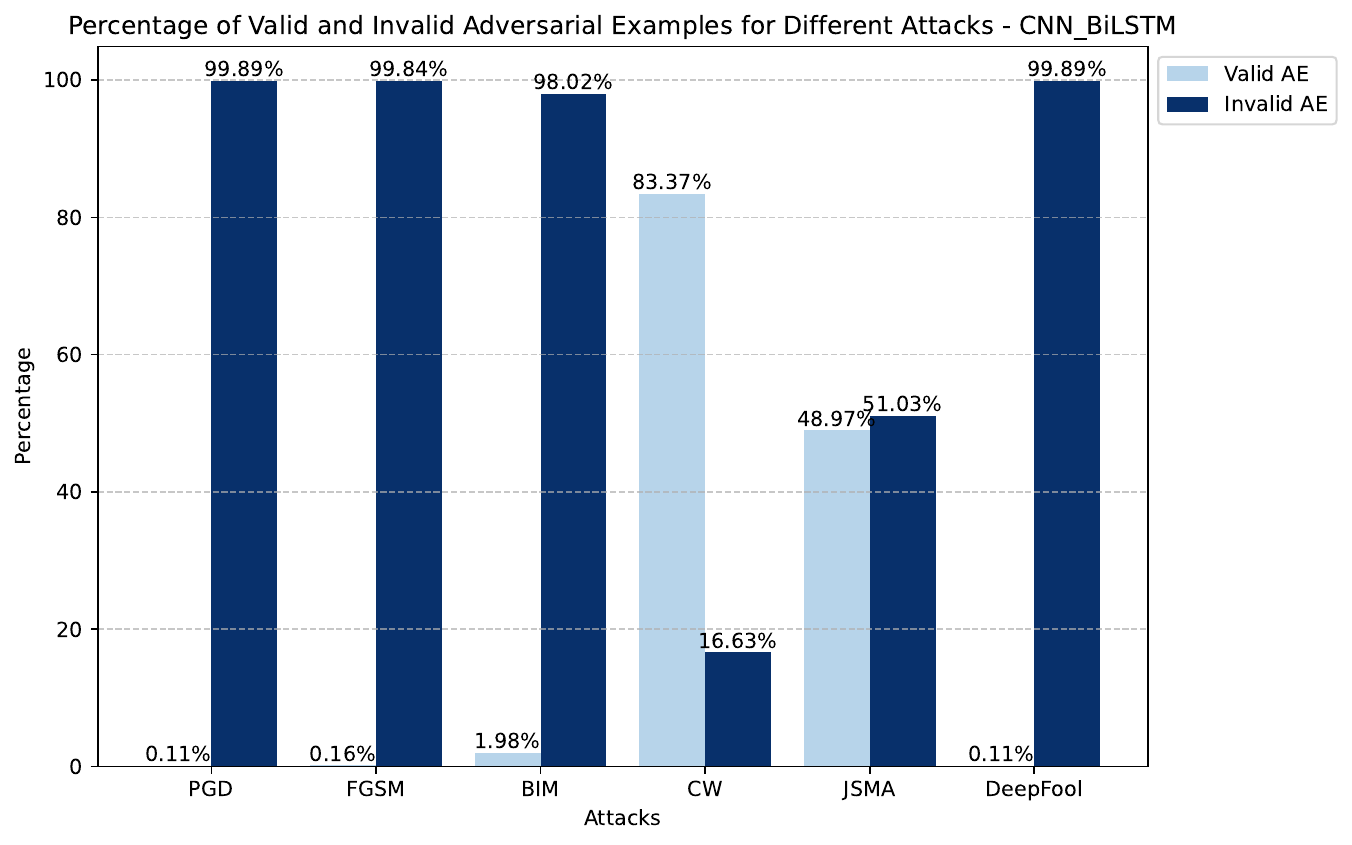}
            \caption[]%
            {{\small CNN-BiLSTM as surrogate model}}    
            \label{fig:mean and std of net44}
        \end{subfigure}
        \caption[ ]
        {\small Effect of different surrogate models on generating valid adversarial examples} 
        \vspace*{-0.4cm}
        
        \label{fig:surrogate}
    \end{figure*}

However, some attack methods remain relatively potent even under these constraints, likely due to their ability to make subtle, targeted changes that respect domain boundaries while still pushing inputs towards decision thresholds.
    
\subsection{Effect of selecting a surrogate model in generating valid adversarial examples}

The experimental results reveal significant variations in the generation of valid AEs across different model architectures and attack methods, as illustrated in Figure \ref{fig:surrogate}. The figure highlights how model complexity influences robustness against adversarial attacks within network domain constraints.

A clear pattern emerges: simpler models like the MLP exhibit a higher proportion of valid AEs across various attacks, including PGD, FGSM, and BIM. In contrast, more complex models, such as CNNs and those incorporating LSTM components, demonstrate increased robustness, with a noticeable reduction in the rate of valid AEs. For certain attacks (e.g., PGD, FGSM, BIM, and DeepFool), the CNN architecture did not yield any valid adversarial examples that complied with network domain constraints. This suggests that more intricate decision boundaries in complex models are more resistant to constrained perturbations, although some attack methods deviate from this trend, underscoring the need to consider both model architecture and specific attack characteristics in adversarial robustness assessments.

The results underscore the importance of selecting appropriate surrogate models in AML, especially when working within domain-specific constraints. More complex architectures could offer enhanced robustness in constrained environments. As the MLP consistently produced more valid AEs across most attacks, it was chosen for the transferability assessment in the subsequent section to evaluate the worst-case scenario regarding vulnerability to adversarial examples.

\subsection{Attacks Transferability}
Two critical factors influence the success of adversarial attack transferability: the inherent vulnerability of the target model and the sophistication of the surrogate model used for attack generation. We assess this transferability by quantifying the variation of accuracy and severity on target classifiers (SVM, Decision Tree, Random Forest, KNN, CNN variants) when exposed to adversarial examples crafted using an MLP surrogate model.  Our findings, presented in Table \ref{table5}, reveal several insights.

\begin{table*}[!t]
\centering
\caption{\small Results of transferability}
\label{table5}

\begin{tabular}[H]{lcccccc||cccccc}
\multicolumn{13}{c}{\textbf{Severity}}\\
\hline
\multicolumn{7}{c}{\textbf{\textit{Before Filter}}} & \multicolumn{6}{c}{\textbf{\textit{After Filter}}}\\
\hline
&JSMA&FGSM&DeepFool&C\&W&PGD&BIM       &JSMA&FGSM&DeepFool&C\&W&PGD&BIM\\
\hline
SVM&36.32&36.09&26.36&35.84&53.80&51.84       &25.55&1.61&1.70&25.55&1.70&1.70\\
DT&36.99&32.56&34.16&32.83&37.96&45.33        &23.29&2.26&2.35&23.29&2.35&2.35\\
RF&41.61&42.11&48.45&39.20&48.03&48.03        &4.27&4.18&4.27&31.39&4.27&4.27\\
KNN&35.39&35.77&38.65&35.29&38.42&37.65       &44.13&0.76&0.85&25.75&0.85&0.85\\
\hline
\hline
CNN&23.83&23.19&41.32&8.53&54.38&54.69        &9.98&0.07&0.08&1.92&0.16&0.16\\
CNN+LSTM&25.84&24.05&39.81&10.14&53.80&53.51  &7.84&0.05&0.04&0.53&0.16&0.15\\
CNN+BiLSTM&23.05&13.73&17.42&10.53&37.92&31.47&5.14&0.04&0.00&0.44&0.16&0.15\\

\hline
\end{tabular}
\end{table*}%

The results indicate a sharp decline in attack severity after filtering, with traditional ML models showing reductions as high as 52.10\% (SVM, PGD) and 44.18\% (RF, DeepFool). However, KNN's increase in JSMA severity (-8.74\%) suggests this attack can still exploit certain vulnerabilities post-filtering. DL models consistently exhibit high resilience, with significant drops, especially in FGSM and BIM, where reductions exceed 50\% across CNN variants. These findings highlight the filtering process's effectiveness, especially for DL models, which appear more capable of neutralizing constrained adversarial attacks compared to traditional models.

\section{Conclusion and perspectives}\label{section:5}

Crucial insights emerge from our investigation into the interplay between domain constraints and adversarial machine learning (AML) within cybersecurity and IoT contexts. Our findings demonstrate that domain constraints significantly reduce the space of potential adversarial examples, enhancing model robustness by hindering attackers' ability to craft deceptive perturbations within valid input ranges. This effect may be partially attributed to deep learning models' inherent capacity to distinguish realistic data points from anomalous inputs, possibly due to implicit regularization during training.

However, our research also exposes limitations in current AML algorithms when applied to cybersecurity, particularly in IoT environments where resource-constrained devices generate distinctive network traffic patterns. Rigorous evaluation within realistic IoT network constraints shows that most existing algorithms struggle to generate feasible and effective real-world adversarial examples. Paradoxically, constraining adversarial examples to valid input ranges often amplifies their impact on model vulnerability, particularly highlighting the effectiveness of black-box transferability attacks in tightly controlled IoT environments.

These results suggest a paradigm shift in AML research for cybersecurity, including IoT-specific applications. Future work must prioritize developing and evaluating defense mechanisms specifically tailored to counter feasible adversarial examples within domain-constrained IoT environments. Additionally, the resilience of targeted, smaller-scale perturbation strategies like JSMA to domain constraints warrants further investigation. To comprehensively address these challenges, researchers could explore diverse datasets and implement proof-of-concept defenses within live IoT network traffic simulations. Such efforts are crucial for building more robust and resilient ML-based intrusion detection systems capable of withstanding sophisticated adversarial attacks while maintaining operational validity within real-world IoT constraints.

% \bibliographystyle{plain}
% \bibliography{paperbib.bib}

\bibliographystyle{IEEEtran}
\bibliography{paperbib}

\end{document}